# Are some books better than others?

**Hannes Rosenbusch**[1] (h.rosenbusch@uva.nl; 0000-0002-4983-3615), **Luke Korthals**[1] (l.korthals@uva.nl; 0009-0006-2098-8679)

[1]University of Amsterdam, Department of Psychological Methods, Amsterdam, Netherlands; Supplementary materials: https://osf.io/46h7s

"Books are well written, or badly written."
- Oscar Wilde

"There is nothing either good or bad, but thinking makes it so."
- Hamlet (William Shakespeare)

**Scholars, awards committees, and laypeople frequently discuss the merit of written works. Literary professionals and journalists differ in how much perspectivism they concede in their book reviews. Here, we quantify how strongly book reviews are determined by the actual book contents vs. idiosyncratic reader tendencies. In our analysis of 624,320 numerical and textual book reviews, we find that the contents of professionally published books are not predictive of a random reader's reading enjoyment. Online reviews of popular fiction and non-fiction books carry up to ten times more information about the reviewer than about the book. For books of a preferred genre, readers might be less likely to give low ratings, but still struggle to converge in their relative assessments. We find that book evaluations generalize more across experienced review writers than casual readers. When discussing specific issues with a book, one review text had poor predictability of issues brought up in another review of the same book. We conclude that extreme perspectivism is a justifiable position when researching literary quality, bestowing literary awards, and designing recommendation systems.**

How much of a novel's merit is debatable? To what degree is good fiction in the eye of the beholder? Newspaper articles and awards committees often provide strong opinions, thereby steering the economic fate of authors and determining what the public considers to be good literature (1). In this paper, we set out to empirically disentangle the relative contributions of book and reader characteristics in determining online book reviews.

**Subjectivity in book ratings**

Whether discussing architecture, music, or written works, people have always debated whether beauty comes down to personal taste or is, at least sometimes, undeniable (2, 3, 4, 5).

In the world of fiction writing, most creators, editors, and publishing professionals agree that different books appeal to different readers (6, 7, 8). As a clear sign of this, people's literary taste changes over their lifespan; the very same reader might love a book as a child but then loathe it a few years later (9). Some books might entertain millions of adoring fans, while being snubbed by professional critics (10). Once celebrated books might lose their appeal to modern audiences (11). All this suggests that a reader's perspective matters when assessing book quality. But how much?

Bizzoni and colleagues discuss whether *mild* perspectivism can provide an adequate description of reading enjoyment, where books do vary in their absolute quality, but differentiable reader groups have their own latent quality score for each book.

As examples of reader groups, the authors list professional critics, lay readers, and demographic groups, which vary in their reading appreciation (12). In fact, readers themselves report enjoying different literary genres (13) and such self-ascribed preferences affect how they search for new books (14), which reading communities they join (15, 16), and which recommendations they receive by librarians and literature websites (17, 16).

The biggest divide in reading enjoyment lies between people claiming that they hate books *in general* and people who make sweeping declarations to love books (18). Long-term reading enthusiasts often connect books to various positive experiences throughout their life from bonding with parents to escaping their sorrows during difficult times (19). They appreciate books as stimulants of their imagination, and often form habits around reading (20). Conversely, book skeptics tend to reject books for reasons of anticipated boredom or exhaustion (21), or because they genuinely don't think they have the skills to read a book (22). Teachers complain that many students lack patience for books or have been poached by fast-paced, visual media (23). Non-surprisingly, a person's overall attitude towards books, as well as their self-perception as a reader, primes their subsequent enjoyment of books (24; 25). In sum, there is widespread agreement that a book's perceived quality depends on characteristics of the reader, at least to some degree.

**Objectivity in book ratings**

While the reader's perspective matters, most people also believe that some texts are truly better than others (26). In fact, books are regularly bestowed with absolutist scores, reviews, and awards that celebrate the "best" literature (27). There are instructional handbooks on how to write better (28), lists of common writing mistakes (29), and detailed how-to-guides by famous authors (30).

Most people would likely agree that a book filled with (unintentional) grammatical errors, logical inconsistencies, and written in a challenging font has some avoidable shortcomings. People are so certain that these things are mistakes that students are taught to avoid them in their writing. Admittedly, professional authors can eradicate such superficial errors quite easily; however, the mere observation that one can get better at writing suggests that authors might still differ on the latent dimension of 'writing skill'. For instance, other common reader complaints, like unlikable or flat characters, are much more abstract and challenging to avoid, and thus more difficult to master than grammatical rule following (31; 32). An analysis by Kaufman and Kaufman suggests that authors often need more than ten years of professional experience to produce their best work (33).

If authors can become better at writing, this entails that books can differ in their quality, provided that an adherence to writing conventions translates to increased appreciation by the readers. Tankard and Hendrickson tested whether one of the most famous writing tips (Show, don't tell!) actually has a positive influence on reading enjoyment (34). They found that the descriptive show-versions of sentences (e.g., "Suddenly I awoke in a drenching sweat, my heart racing") were indeed perceived as more interesting and engaging than the abstract tell-versions ("Suddenly I awoke, frightened"). However, not all writing conventions correlate with book popularity as Boyd and colleagues found no benefit of choosing any narrative structure over another in a sample of 60,000 works of fiction (35).

In sum, it is a common belief, backed by some empirical evidence, that a story can be written well or poorly, meaning that some books might truly be better than others. We are left with two questions. First, to what *degree* is a book's perceived quality due to the book, versus the reader? And second, how could one measure a book's quality?

**How to measure book quality**

It turns out that the aforementioned questions are two sides of the same coin. Both, the nature and the measurement of book quality, pose the problem of which reviewer to believe. One field that battles this problem is computational literary studies, where machine learning algorithms are employed to predict a book's success.[1]

While some researchers focus on predicting book sales or download counts (36, 37, 38, 39, 40, 41), others explicitly forecast canonicity, award wins, or review scores (42, 43, 44, 45). Such prediction models have the power to steer publishers' investments and therefore determine what the public reads and which authors become successful. However, can we assume that the chosen prediction targets–specifically reader reviews and committee scores–are reliable measures of a book's enjoyability?

Bizzoni and colleagues (12) describe two opposing views in this regard: strong perspectivism and weak perspectivism. The former denotes that even a single review score constitutes an accurate measurement of a book's quality, but might not generalize to other readers. Thus, a book's quality should only be measured and predicted for the individual reader, and never in absolute terms. Weak perspectivism, on the other hand, argues that book quality *can* be measured in absolute terms, but might require averaging across many reviews by different people. It postulates that individual review scores serve as noisy indicators of a book's latent quality. However, *how much* idiosyncratic noise overlays the latent book quality remains unknown.

Researchers predicting aesthetics ratings in other fields have quantified the amount of inter-individual variation to distinguish it from other forms of statistical 'noise'. For instance, Hönekopp found that judgments of facial beauty are about equally determined by face-characteristics and rater-tendencies (46). Hehman and colleagues extended this finding by showing that some face ratings truly depend on the observed face (e.g., perceived happiness) whereas other traits are almost entirely due to the beholder (e.g., perceived creativity; 47).

Human-generated art like paintings, architecture, and textiles elicit more person-specific aesthetics ratings than faces and nature scenes (48), a trend that is more pronounced for abstract art than representational art (49; 50).

When it comes to text evaluations, there is evidence for both subjectivity and objectivity. The quality of news domains, for instance, appears to be rated with high agreement among experts (51), whereas reviewers of scientific manuscripts often disagree in their final evaluations (52).

The appreciation of textual humor appears to be largely subjective (53) and, as textual humor is a form of creative writing, we decided to pre-register our numerical hypotheses—about the relative importance of books and readers—in direct accordance with the confidence intervals for written jokes (cf. 53; study 2):

*Relative-importance-hypothesis*
Differences between raters will account for more than three times as much rating variance as differences between books.

---

[1] Notice the clichéd 'coin' expression in the section's first sentence. Did it hamper your reading enjoyment (as it is unoriginal), or did you find it useful (as it is succinct)?

*Book-importance-hypothesis*
Differences between books will account for five to nine percent of all rating variance.

*Rater-importance-hypothesis*
Differences between raters will account for thirty to thirty-eight percent of all rating variance.

Next to broadening our knowledge of subjectivity in text evaluations, the presented analyses will provide answers to practical questions like: Should one use review scores for purchasing decisions? How many reviewers are needed to assess a book's quality and make fair comparisons between books? And which pieces of information are needed to predict a book's reception by the public?

**Materials and Methods**
Pre-registration, code, and data can be found here: https://osf.io/46h7s. We analyze book reviews from one of the largest literary websites: goodreads.com (cf., 32). Goodreads users rate books by clicking on a star scale, reaching from one to five stars, or by writing a free text review.

To collect the data, a Python script iterated through random numbers and inserted them into the website's url, stopping after 300,000 attempts. From each random user, up to twenty random reviews were collected. Private accounts, and reviews that weren't posted publicly for the entire reading community, were not accessed. Users with zero or one review were also discarded. No identifying data were collected. The final dataset consists of 566,121 ratings, from 67,012 participants, and across 49,674 books. Users annotated the book corpus with 875 unique genres ranging from very general ("fiction") to very specific tags ("World War II").

During the analyses, we disentangle two sources of variance, the book and the reviewer, through the use of random-effects in multilevel models and the ICC value associated with each source of variance. We further show probabilities of book ratings conditional on other ratings to assess rating consistency across readers and books. Confidence intervals are computed by bootstrap sampling each analysis with ten repetitions and noting the highest and lowest values. Note that we pre-registered 100 rather than ten repetitions, which were unneeded given the large sample size.

**Results**
A multilevel model with random intercepts per book and reader (54) revealed that differences between books account for 3.53% [3.42–3.68] percent of the rating variance on Goodreads, whereas differences between raters accounted for 30.28% [30.06–30.51]. Reader differences were 8.58 times [8.2–8.87] more impactful than differences between books. Figure 1 illustrates the relative benefit of knowing the rater vs knowing the book when trying to predict rating scores.

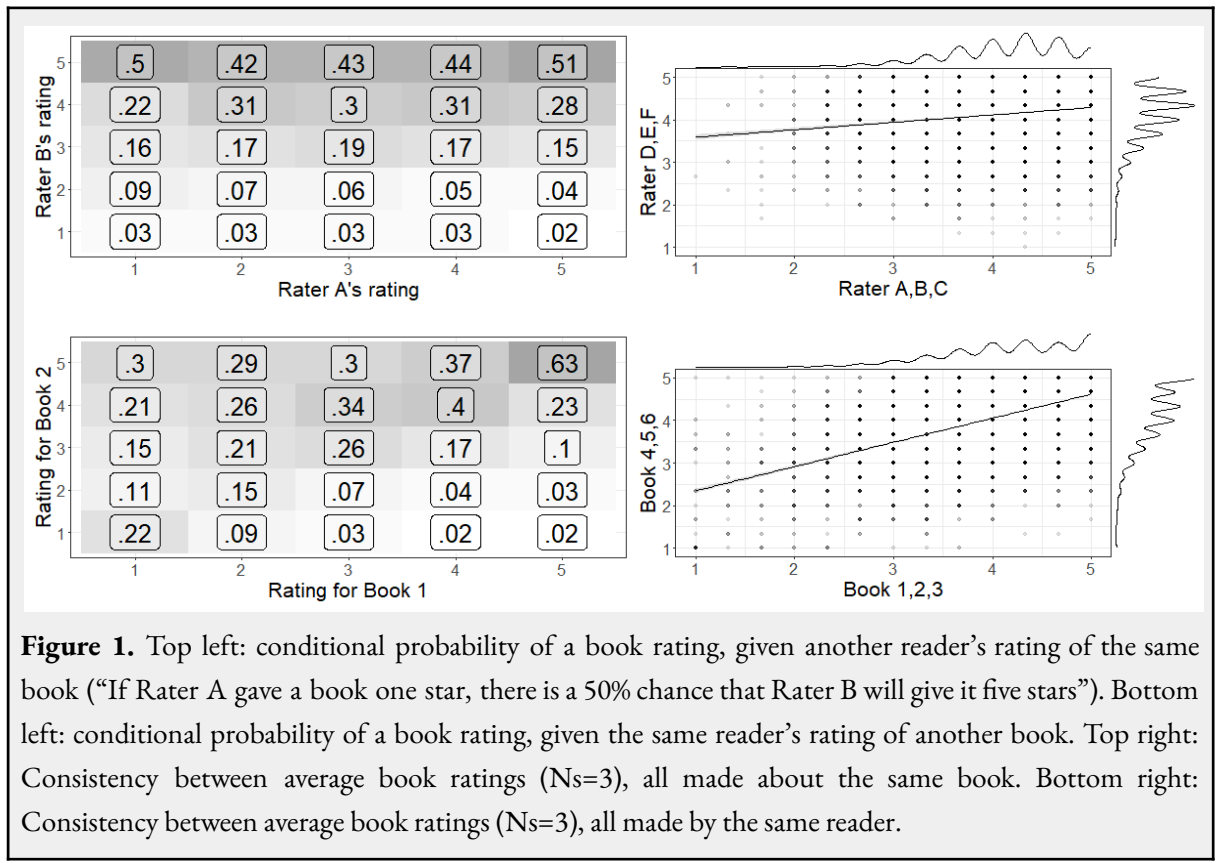

**Figure 1.** Top left: conditional probability of a book rating, given another reader's rating of the same book ("If Rater A gave a book one star, there is a 50% chance that Rater B will give it five stars"). Bottom left: conditional probability of a book rating, given the same reader's rating of another book. Top right: Consistency between average book ratings (Ns=3), all made about the same book. Bottom right: Consistency between average book ratings (Ns=3), all made by the same reader.

The relatively high consistency of ratings from a given reader, compared to the lack of consistency for ratings of the same book, is also highlighted by an extended simulation where we sampled an increasing number of raters and averaged their score to approximate either a single rater's score, or the book's average score in our complete dataset (cf. Figure 2).

**Subsample analyses**

When restricting the book sample to written fiction books (i.e., excluding non-fiction, audiobooks, and image content; N = 26,160), the variances accounted for by book differences (3.23% [3.07-3.39]) and rater differences (29.61% [29.26-29.92]) stayed virtually the same. Another sensitivity analysis on the 50% of books with the least amount of ratings, showed a similar impact of books (3.23%; [2.52-4.03]) and an even higher impact of the rater (40.76%; [40.35-41.14]). When removing all users that gave the same rating to all their reviewed books (which includes users with few ratings, user that seemingly use the platform to collect favorite books, and potentially undiscerning book enthusiasts), book differences accounted for (4.3%; [4.06-4.47]), while reader tendencies by definition diminished in importance but were still considerably higher (23.14%; [22.76-23.49]; $N_{ratings}$ = 497,413).

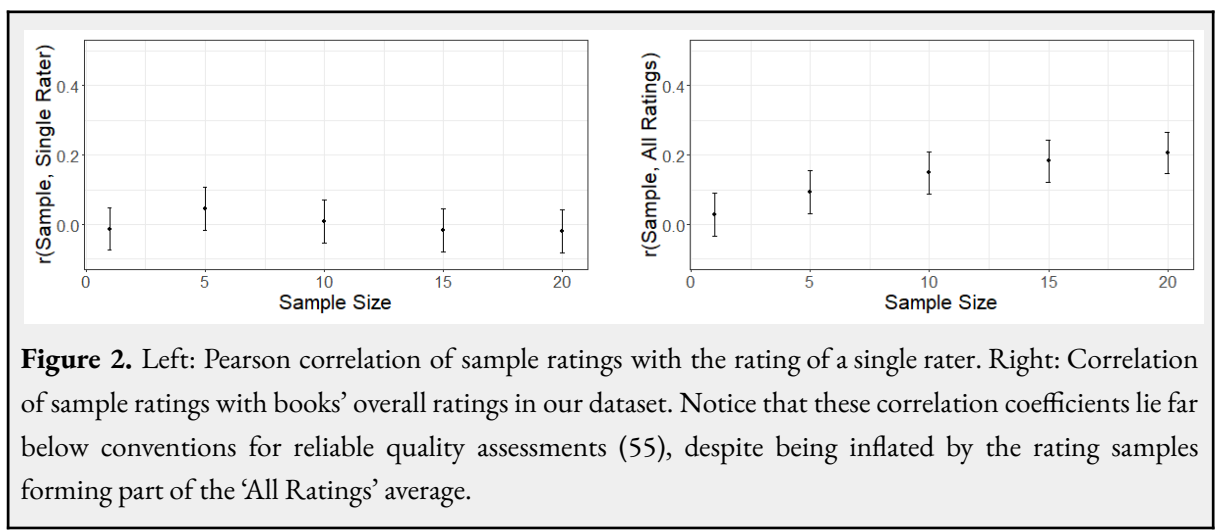

**Figure 2.** Left: Pearson correlation of sample ratings with the rating of a single rater. Right: Correlation of sample ratings with books' overall ratings in our dataset. Notice that these correlation coefficients lie far below conventions for reliable quality assessments (55), despite being inflated by the rating samples forming part of the 'All Ratings' average.

It is possible that the amount of rater agreement is higher within genres, as 'genre tourists' might introduce reader-level variance through a less practiced eye for genre-adequate writing. Thus, we restricted the sample to book ratings from readers who had rated books from the same genre positively (4 or 5 stars) at least twice[2]. The resulting dataset of 433,149 'within-genre-ratings' showed that book differences still accounted for fairly little variance (3.86%; [3.68-4.09]), while the importance of rater tendencies diminished (22.63%; [22.37-22.97]), potentially due to overall higher ratings. Lastly, it is conceivable that readers who are practiced in reading and reviewing books achieve a higher agreement in their evaluations. In line with that reasoning, we find that users who rated more than ten books on Goodreads and additionally wrote more than five freetext reviews (N = 5,464), showed less idiosyncratic variance (18.19% [17.33-19.05]) than the full sample analyzed above. Further, book contents accounted for about twice as much variance in the expert sample compared to the full sample (7.67% [7.31-8.25]), although still less than idiosyncratic rater tendencies by a factor of 2.37 ([2.16-2.58]; $N_{ratings}$ = 54,016).

---

[2] We defined 'same genre' books as having at least three shared tags out of the twenty most common genre tags, or the exact same set of genre tags (in case the book had less than three tags).

**Analysis of freetext reviews**

The analysis of practiced reviewers indicates that book contents *can* account for at least some rating variance under certain circumstances. However, most of the time, this latent consensus appears to be buried under masses of casual, idiosyncratic evaluations. In that context, it is worthwhile pointing out that Goodreads users do not have to provide proof that they have purchased or read a book before rating it. Thus, our full sample likely includes ratings made for self-presentational purposes, ratings submitted without having read the book, and ratings submitted after a mere skimming of the book contents. Such superficial ratings could have inflated the importance of idiosyncratic reader characteristics and wash out the importance of book contents in our variance decomposition. Thus, we replicated the analyses above, but focusing on people's freetext reviews (N=58,199) rather than their star ratings. We assume that written reviews include more reliable information about people's reading experience as the writing process requires deliberate introspection, while likely decreasing the rate of fraudulent reviews (the large majority were written 'pre-gpt').

In order to produce the numerical values required for variance decomposition we condensed each review into a binary sentiment score, using a distilbert model (56, 57). Given the binary scale of the outcome variable, we updated the previous multilevel model with a logistic link function. Logistic models do not provide the residual variance term required for the computation of ICCs. Thus, we used the latent threshold technique (58) under which the binary outcome variable is assumed to have an underlying continuous scale with a logistic error distribution. In our case the assumption of a latent continuum is warranted given that review sentiments are not required (and in fact unlikely) to be binary by nature. Further, the assumption of a logistic error variance does not affect the *relative* importance of reviewers and books, which came down to a factor of 3.36 [3.08–3.65] in favor of reviewer tendencies. The estimated absolute ICC values ascribed 13.85% [13.22–14.66] of the review variance to differences between reviewers and 4.14% [3.65–4.48] to differences between books.

**Analysis of review topics**

As a final analysis, we examined whether readers bring up similar issues when reviewing the same book. It might be that readers disagree widely in their overall book ratings, while still perceiving a similar set of strengths and weaknesses per book (of which the importance could vary idiosyncratically). We used a large language model (GPT-4o) to annotate written reviews ($N$ = 26,699) for fiction books regarding the mentioning of feeling bored, addicted, or confused by the book contents, as well as the mentioning of characters, and the author's unique style or skill in writing. We chose these concepts as they relate to common facets of enjoyment and broad book distinctions (cf., page-turning thrillers vs literary/character-driven fiction). We also annotated the presence of a book summary in the review as a control variable, as we didn't expect books to differ much in whether they elicit summaries from readers. GPT4o achieved an annotation accuracy of 98.6% [97.74-99.11] on 200 human-annotated book reviews (for details, see supplementary materials).

For our analysis, we drew pairs of reviews either targeting the same book or stemming from the same reviewer. As shown in Figure 3, the presence of specific review attributes was better predicted by other reviews of the same reviewer, compared to other reviews about the same book. Especially the writing of summaries seems to be highly characteristic of specific reviewers.

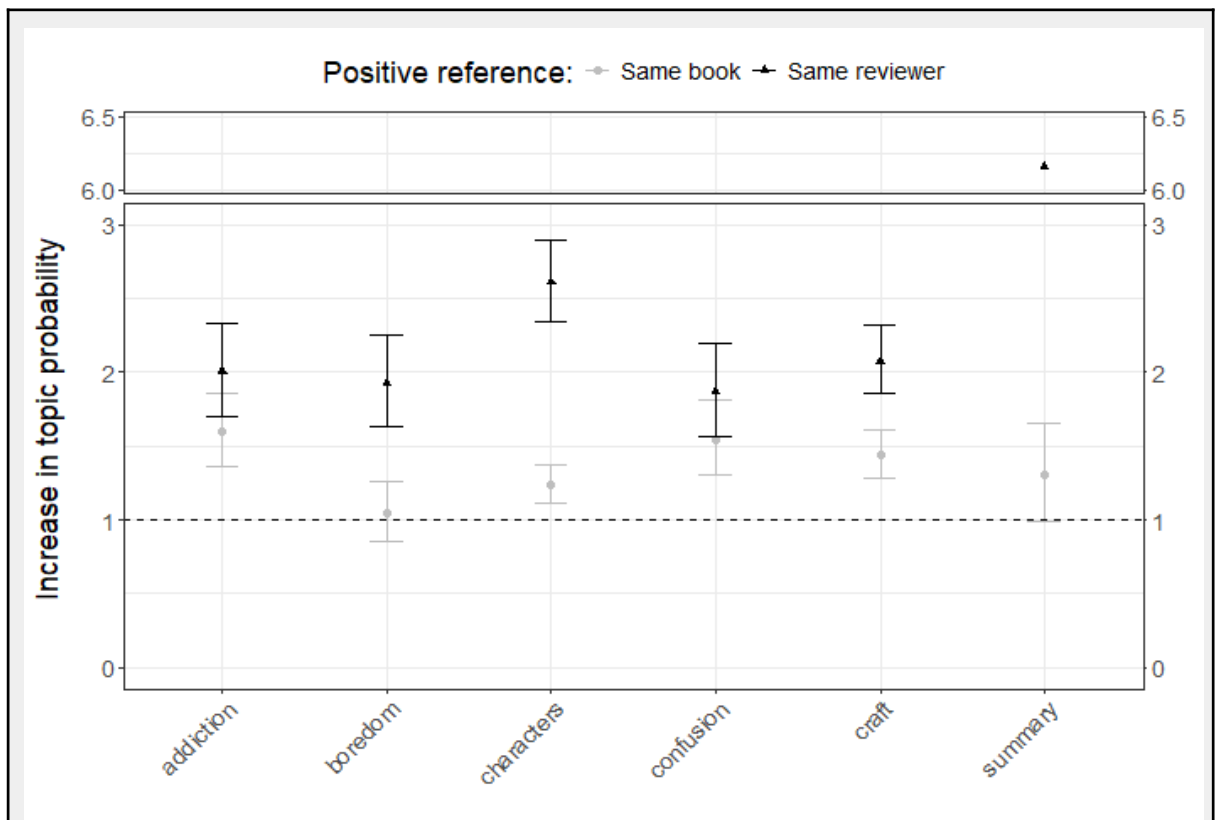

**Figure 3.** The X axis shows various topics that were brought up in book reviews. The Y axis shows how much more likely it is that a review mentions this topic if it was mentioned in a reference review (i.e., $\frac{Probability(Topic \mid mentioned\ in\ reference)}{Probability(Topic \mid not\ mentioned\ in\ reference)}$. For instance, the black triangular point on the left indicates that a review is twice as likely to mention that the reviewer felt addicted to the book, if the *same reviewer* made that claim about a different (randomly chosen) book. The gray circle below the triangle indicates that a review is about 1.5 times more likely to contain 'feeling addicted' if another (randomly chosen) reviewer made that statement about the *same book*.

**Discussion**

How much value should you place on a stranger's opinion when looking for an enjoyable read? Our analyses suggest that you shouldn't place any at all. A book evaluation will generally tell you much more about the reviewer than the book–up to ten times more in our analyses.

While it is tempting to proclaim aesthetic value in absolute terms ('must-reads' vs 'trash-novels'), we find rater characteristics had a much stronger impact on book reviews than book contents. This is in line with previous work on the appreciation of jokes and abstract art (53, 49). Across both star ratings and written reviews, the impact of book characteristics (3-4% of rating variance) was even lower than our a-priori estimates (5-8%).

While genre fans also differed widely in their evaluations of genre books, experienced reviewers achieved higher agreement (7-8%). That is to say that they managed to approximate the average book evaluation better than casual readers. However, before trusting the review of an expert, one should consider that even an expert's approximation remains virtually uncorrelated to the reading enjoyment of another individual reader. All in all, we conclude that one shouldn't just apply mild but extreme perspectivism when talking about the enjoyability of books.

In addition, we found that even specific complaints and compliments in written book reviews are more characteristic of the reader than the book. Thus, we speculate that readers don't just apply a different set of weights for consensual book criteria, but rather apply altogether different criteria (59, 60). People read books for different reasons and thus end up bestowing different evaluations and critiques (60). Thus, it is all the more noteworthy when reviewers *do* agree in their comments, as such an outlying degree of consensus hints at book contents that deviate so much from the norm that readers can't help but converge in their reviews. For instance, today (30th of September 2024) the latest reviews of the book "If on a Winter's Night a Traveler" almost all mention that it is confusing or at least challenging to read (which some readers report to enjoy), indicating that unique books lead to more homogenous review contents.

In the world of scientific peer review, it is well-known that manuscript evaluations are rarely consensual and often deviate from some 'true' metric (55, 52). A popular method to strengthen the association between text contents and evaluation is the introduction of narrowly defined rubrics (61). For fiction writing, such rubrics could target criteria which are believed to predict reading enjoyment for a wide audience (e.g., "Does this book feature a protagonist with desirable characteristic X?" or "Is the adverb prevalence in this book below X%?"). While high-dimensional annotations of books can likely be generated by AI models (cf., our annotation of book reviews above; 62) it remains to be seen whether such rubric-style annotations can approximate a single person's reading enjoyment, especially as reading preferences clearly vary across, for instance, people's motivations and personality traits (59, 60).

Viewing the current results, people certainly shouldn't feel discouraged to read reviews or rate books themselves. Such activities are likely to enhance one's reading appreciation by adding social relevance and offering new perspectives on books (63).

While absolut book rankings can serve as indicators of prestige (see 64 for a review of quality metrics), they cannot predict a person's enjoyment of a book. However, rather than eradicate them, they could simply be seen as descriptive of current audience trends or marketing success. Themed lists and book awards also offer value by directing readers towards specific book contents that they are interested in.

In that regard, it is also noteworthy that the lacking 'accuracy' of book evaluations does not necessarily preclude accurate book recommendations. When a reader's relevant book preferences are measured accurately and matching books can be identified reliably, experts or recommendation engines can still function (as long as the measured preferences are sufficiently stable and highly predictive of reading enjoyment). However, general ratings and rankings can be largely discarded for recommendations and numerical scores of individual readers can be fully discarded, at least for professionally published books.

**References**

1. Chuey, A., Luo, Y., & Markman, E. M. (2024). Epistemic language in news headlines shapes readers' perceptions of objectivity. *Proceedings of the National Academy of Sciences of the United States of America*, *121*, e2314091121.

2. Ball, P. (2004). Measuring beauty. *Nature*, *1476*.

3. Hyman, J. (2002). Is beauty in the eye of the beholder? *Think*, *1*, 81–92.